# Quantum well states in two-dimensional gold clusters on MgO thin films

X. Lin, N. Nilius,[*] H.-J. Freund

*Fritz-Haber Institut der MPG, Faradayweg 4-6, D14195 Berlin, Germany*

M. Walter,[†] P. Frondelius, K. Honkala, H. Häkkinen[*]

*Departments of Physics and Chemistry, Nanoscience Center, University of Jyväskylä,*
*P.O. Box 35 (YFL), FI-40014 Finland*

The electronic structure of ultra-small Au clusters on thin MgO/Ag(001) films has been analyzed by scanning tunneling spectroscopy and density functional theory. The clusters exhibit two-dimensional (2D) quantum well states, whose shapes resemble the eigen-states of a 2D electron gas confined in a parabolic potential. From the symmetries of the HOMO and LUMO of a particular cluster, its electron filling and charge state is determined. In accordance to a DFT Bader-charge analysis, aggregates containing up to twenty atoms accumulate one to four extra electrons due to a charge transfer from the MgO/Ag interface. The HOMO – LUMO gap is found to close for clusters containing between 70 and 100 atoms.

[*] Corresponding authors: nilius@fhi-berlin.mpg.de,  Hannu.J.Hakkinen@jyu.fi
[†] Present address:  Fakultät für Physik, Universität Freiburg, S. Meier Str. 21, D79104 Freiburg, Germany

Ultra-small metal clusters exhibit unique size-dependent properties due to their large fraction of surface atoms and distinct electronic structure.[1] Reducing the atom count below a few hundred, the bulk band structure breaks down and discrete energy levels separated by gaps emerge. These quantization effects, in conjunction with geometric particularities, are responsible for the unusual chemical properties of small metal particles as well as for a variety of optical and electronic phenomena.[2,3] For example, oxide-supported Au clusters show a pronounced reactivity maximum in the low-temperature CO oxidation in the size range of the non-metal to metal transition.[4,5,6]

In spite of this importance, electron quantization effects are difficult to access experimentally. So far, the most comprehensive picture has been deduced from photo-electron spectroscopy on size-selected clusters in the gas-phase.[7,8] Such experiments cannot account for substrate effects on the cluster electronic structure, which might however decisively influence the properties of particle-support systems.[5,6,9] In contrast, the investigation of ad-clusters is often hampered by their size and shape distribution on the surface, giving rise to inhomogeneous broadening effects in non-local spectroscopies. Spatially resolving techniques, e.g. scanning tunneling microscopy (STM), provide a powerful alternative to study the interplay between geometric and electronic properties at the single-particle level.[10,11,12]

This paper reports on a combined STM and density functional theory (DFT) study of ultra-small Au clusters grown on MgO thin films on Ag(001). The Au-MgO system belongs to the best-explored metal-oxide system, thanks to the possibility to prepare high-quality oxide films[13,14] and the simple MgO rock-salt structure that facilitates theoretical modeling. Nano-dispersed gold follows a specific adsorption scheme on MgO films that is governed by a charge transfer from the metal support into low-lying Au affinity levels.[15,16,17] An $Au^-$ species strongly interacts with the MgO surface, as it locks into the oxide Madelung field, induces a polaronic lattice distortion and polarizes the metal support underneath.[15,17] Also the two-dimensional (2D) growth regime of larger Au clusters on MgO thin films,[18] being in contrast to the 3D growth on bulk MgO, has been attributed to electron transfer processes.[19,20] However, experimental evidence for this charge exchange has been obtained only for linear Au clusters on alumina so far.[21] This work now provides detailed insight into the level quantization and the charge state of individual Au clusters deposited on a 2 ML thin MgO film.

The experiments are carried out with an ultra-high-vacuum STM operated at 4.5 K. The sample electronic structure is probed with differential conductance (dI/dV) spectroscopy using lock-in technique ($U_{mod}$ = 10 mV). The MgO film is prepared by reactive Mg deposition onto a sputtered/annealed Ag(001) surface in $1\times10^{-6}$ mbar $O_2$ and 570 K. The procedure results in an atomically flat MgO film, exposing large rectangular terraces delimited by non-polar step edges.[13,16] Single gold atoms are evaporated from a high-purity Au wire and deposited onto the sample at 100 K. Due to the thermal energy of the incoming atoms and the low diffusion barrier, Au aggregates into ultra-small clusters on the oxide surface. The DFT calculations are performed with the Grid-Projector-Augmented-Wave method (0.2 Å grid-spacing)[22] using the generalized gradient approximation.[23,24] Periodic boundary conditions are applied in the surface plane, whereas the perpendicular direction is kept non-periodic. The substrate is modeled with three Ag(001) and two MgO layers on top, using the theoretical Ag lattice parameter of 4.14 Å. The vacuum gap above and below the slab amounts to 6 and 4 Å, respectively. Except for the two bottom Ag layers, all atoms are relaxed during geometry optimization until the forces are below 0.05 eV/Å. Depending on the cluster size, (4×4), (5×5) and (6×6) surface cells are constructed. A Monkhorst-Pack 2×2×1 k-point sampling and the Bader method is used to analyze the electronic structure and the charge state of the ad-clusters. The STM images are simulated with the Tersoff-Haman formalism,[25] using a state density of $1.4\times10^{-5}$ e/Å$^3$ to match the experimental tunnel current of 5 pA.[26]

Fig.1A shows a typical STM topographic image of a 2 ML MgO film after deposition of 0.02 ML Au. The majority species on the surface are round protrusions of 0.8 Å height and 6-7 Å diameter, which are assigned to Au monomers. In addition, a variety of small aggregates with distinct 1D and 2D shapes are observed.[18,27] They can be divided into 'low' and 'high' clusters, according to their apparent height in STM images taken in the accessible bias window (-1.3 to +1.3 V).[28] 'Low' clusters have bias-independent heights of 0.8-0.95 Å, compatible with single-layered Au aggregates, whereas 'high' ones are almost 50% taller (Fig.1B). This height increase does not indicate the presence of a second atomic layer, but results from the availability of electronic states in the cluster that facilitates the electron transport between tip and sample. This becomes apparent in dI/dV spectra of 'high' clusters, which exhibit peaks below and above the Fermi level ($E_F$) corresponding to the highest occupied (HOMO) and lowest unoccupied orbital (LUMO) (Fig.2). The HOMO-LUMO gap was found to vary between 0.1 and 1.8 eV as a function of cluster size.

Apart from this dI/dV fingerprint, 'high' clusters undergo distinct changes in their topographic appearance as a function of imaging bias. Inside the HOMO-LUMO gap, the clusters exhibit compact shapes that resemble the ones of the 'low' species and reflect their topographic structure. Close to the dI/dV peak positions, 'high' clusters are imaged as flowerlike protrusions, indicating that the contrast is now governed by their electronic properties (Fig.1C).[20,25] The particular orbital that is responsible for the modified appearance is identified with dI/dV imaging performed at the respective peak energy. In most cases, the orbitals are characterized by an irregular lobe pattern, manifesting the disordered arrangement of Au atoms within the aggregate (Fig.1C). In selected cases, also highly symmetric orbitals with well-defined nodal planes are observed, which closely resemble the quantum well states (QWS) of a free-electron gas confined in a 2D parabolic potential. The presence of such states with *S,P,D,F* and *G*-symmetry (corresponding to an angular momentum quantum number of 0 to 4) has been predicted for symmetric Au clusters on thin film and bulk MgO.[20] The surprising simplicity of the QWS is owed to the dominance of *Au 6s*-derived states within the MgO band gap (6.4 eV for a 2 ML film)[13] that are well separated from the lower-lying *Au 5d* band.

Based on the distinct HOMO and LUMO shapes and the size information from the STM images, the true atomic and electronic structure of such Au clusters can be determined by comparing the DFT signature of possible candidate clusters with the experimental results. The procedure is demonstrated for the cluster in Fig.2. In low-bias images, it has pentagonal shape and a diameter of roughly 14 Å. At bias voltages above +0.7 V and below -0.3 V, flowerlike patterns with eight lobes and four nodal planes become visible, indicating a *G*-symmetry of both frontier orbitals. Conductance measurements reveal an even more complex picture. In fact, only the HOMO at -0.4 V and the LUMO at +0.8 V are of *G*-character, whereas two additional maxima at -0.8 V (HOMO-1) and -1.2 V (HOMO-2) show *P*-symmetry with the node being orthogonal in both cases. The shape of the HOMO-1 slightly interferes with the HOMO symmetry, which still controls the integral conductance of the junction and therefore the tip-sample distance. Consequently, additional maxima appear in the dI/dV maps of the HOMO-1 at the node positions of the HOMO. The same effect is observed for the LUMO+1, where artificial dI/dV features become visible at the nodal planes of the LUMO. In all images, small differences between the left and right half of the cluster are apparent, indicating a slight asymmetry in its atomic configuration.

Based on this information, an extensive structural search for monolayer Au clusters of 6 to 24 atoms with matching properties was conducted. A planar $Au_{18}$ cluster with $D_{2h}$ symmetry perfectly reproduces the experimental observations (Fig.3A). The structure is derived from a $D_{6h}$-symmetric $Au_{19}$ cluster with one corner atom removed, giving rise to the pentagonal shape observed in STM.[24] The $Au_{18}$ cluster has eleven occupied valence states, ten of them filling the first four electronic shells of the harmonic potential (Fig.3D). The HOMO and LUMO therefore correspond to the lowest and second lowest orbital of the $5^{th}$ shell and are both of *1G*-character. The slight asymmetry in the atomic configuration introduces an energy difference between both states and determines the HOMO-LUMO gap. It is also responsible for the somewhat reduced state density of the LUMO at the position of the missing atom. Both results, the *G*-character of the frontier orbitals and the asymmetric LUMO perfectly matches the experimental observations. The agreement is further supported by the calculated symmetry of the HOMO-1 and HOMO-2 of the $Au_{18}$, which are the two orthogonal *2P*-QWS states in the fourth electronic shell (Fig.3D). The single nodal plane of the *P*-orbital is clearly visible in the dI/dV images taken at –0.8 and –1.2 eV (Fig.2B). From the complete filling of eleven QWS, the number of valence electrons in the cluster is determined to be 22. Each of the 18 Au atoms donates its *6s* electron to the delocalized QWS, implying that the missing four electrons originate from the support. The presence of four excess charges is corroborated by a Bader analysis that yields a value of -3.54e for the $Au_{18}$, in agreement with an average transfer of 0.2e per atom calculated for a compact Au layer on 2 ML MgO/Ag(001). The distinct orbital structure of $Au_{18}$ therefore provides clear evidence for the negative charging of the ad-cluster and enables the first reliable quantification of the underlying electron transfer from the MgO/Ag interface.[15,16,17,19]

The concept of reconstructing the true atomic and electronic structure from the STM data has been applied to other Au clusters on the MgO film, aiming for more general conclusions on the size-dependent cluster properties. Fig.3B shows a slightly smaller aggregate, where the HOMO is of *F*-symmetry and the LUMO, 1.0 eV higher in energy, has *P*-character. Comparing these findings to DFT results of various candidate clusters, best matching is achieved for a $D_{4h}$-symmetric $Au_{14}$. The Bader analysis reveals that this cluster carries two transfer electrons from the support, bringing the total electron count to 16. This charge state is compatible with the level occupancy predicted by the harmonic oscillator model. The HOMO is the 8$^{th}$ QWS and has *1F*-character, while the LUMO is of *2P*-symmetry, similar to the experimental observations. The smallest cluster with well-defined QWS in this study has frontier orbitals of *D*- and mixed S,P-character, separated by an energy gap of 1.9 eV (Fig.3C). The underlying configuration is assigned to a two-fold negatively charged $Au_8$ cluster, whose HOMO and LUMO are the *1D* and *2S*-QWS in the harmonic potential, respectively. While the calculated HOMO matches perfectly, the LUMO symmetry slightly differs from the experiment. The deviations might be explained by a certain overlap between the LUMO and the *2P*-like LUMO+1, which emphasizes the two outer lobes in the measurement. For all Au clusters, QWS with higher angular momentum in a distinct shell have found to be lower in energies than their *S* and *P*-like counterparts, in contrast to the ideal oscillator model. This finding indicates a certain admixture of box character to the harmonic potential that introduces a downshift of states with higher angular momentum.[8]

Finally, the evolution of the HOMO-LUMO gap ($E_g$) is analyzed for clusters of different area $\Omega$ (number of atoms) on the MgO film (Fig.4A). Spectroscopy reveals a gradual closing of the gap around $E_F$ until metallic behavior sets in at around 80 atoms per cluster.[4,6] The gap size follows the inverse cluster area according to $E_g \propto \Omega^{-1}$ (dashed line), as expected for the energy separation of eigenstates in a 2D harmonic or spherical box potential.[8] The smooth dependence of the gap size on $\Omega$ is superimposed by oscillations that reach 0.8 eV for ultra-small clusters. Large $E_g$ values hereby suggest electronic shell closing, whereas smaller gap sizes are compatible with open-shell configurations. The clear oscillatory pattern predicted for a harmonic oscillator[20] is, however, blurred by the presence of asymmetric Au clusters on the MgO, where the degeneracy of QWS is lifted and the gap size is enlarged (Jahn-Teller effect). Furthermore, energetically unfavorable clusters might be removed from the ensemble, e.g. by rearranging their internal atom configuration, attaching further atoms, or adapting the charge transfer from the support to enforce shell closing. On the other hand, the electron transfer into the Au clusters is with 0.2e/atom nearly proportional to the atom number in the investigated size range (Fig. 4B). Interestingly, the accumulated charge is lower in 2D than in 1D clusters for small atom counts, which manifests a reduced internal Coulomb repulsion in linear structures due to the better delocalization of the extra electrons.[11,27] This effect diminishes for clusters containing more than 7-8 atoms, where the 2D growth of Au becomes energetically favorable.[21,27]

In conclusion, distinct QWS in 2D Au clusters grown on MgO/Ag(001) thin films have been imaged with STM and interpreted by extensive DFT calculations. The QWS behavior originates from the confinement of the quasi-free *Au 6s* electrons in the ultra-small aggregates. For sufficiently symmetric clusters, the QWS have orbital shapes similar to those of a 2D electron gas confined in a parabolic potential. In this case, the symmetry of the HOMO and LUMO is exploited to deduce the atomic configuration, the electron filling and the charge state of the clusters via DFT calculations. With increasing cluster size, a linear accumulation of excess electrons is revealed, verifying the long-debated charge transfer mechanism on the MgO/Ag support. This STM-DFT study opens new insights into the interplay between the geometric and electronic structure of individual oxide-supported metal clusters, being an important step towards the understanding of the unique chemical properties of such material systems.

**Acknowledgments** This work is supported by the A.v. Humboldt foundation (X.L.), the Academy of Finland (M.W., P.F., K.H., H.H.), the DFG (M. W.) and the COST D41 network. We thank CSC (Espoo, Finland), FZ Jülich, and EU DEISA for providing the computer resources.

**Figures**

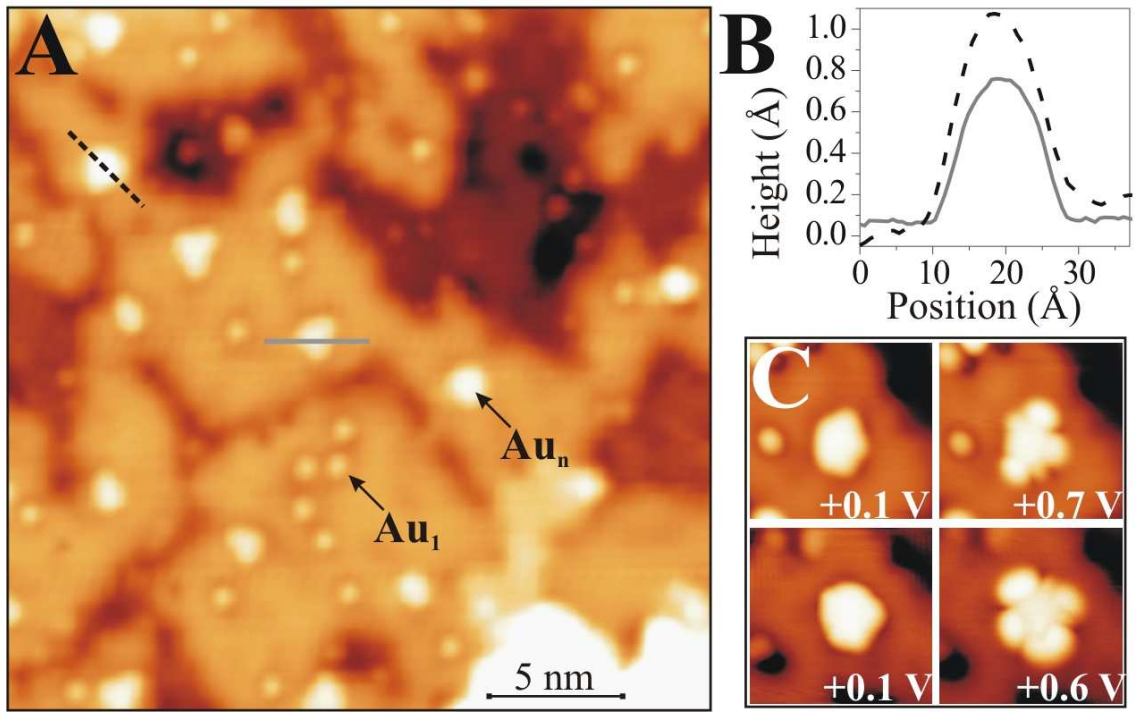

**Fig. 1**
(**A**) STM image of Au adatoms and clusters on 2 ML MgO/Ag(001) ($U_s$ = 0.4 V, I = 5pA, 26×26 nm$^2$). (**B**) Line profiles of the two clusters depicted by the solid and dashed line in (A). The first cluster appears lower due to the absence of QWS below the scan bias. (**C**) Topographic images of an asymmetric and symmetric Au cluster taken in the HOMO-LUMO gap (left) and above the LUMO position (right) (5.5 × 5.5 nm$^2$).

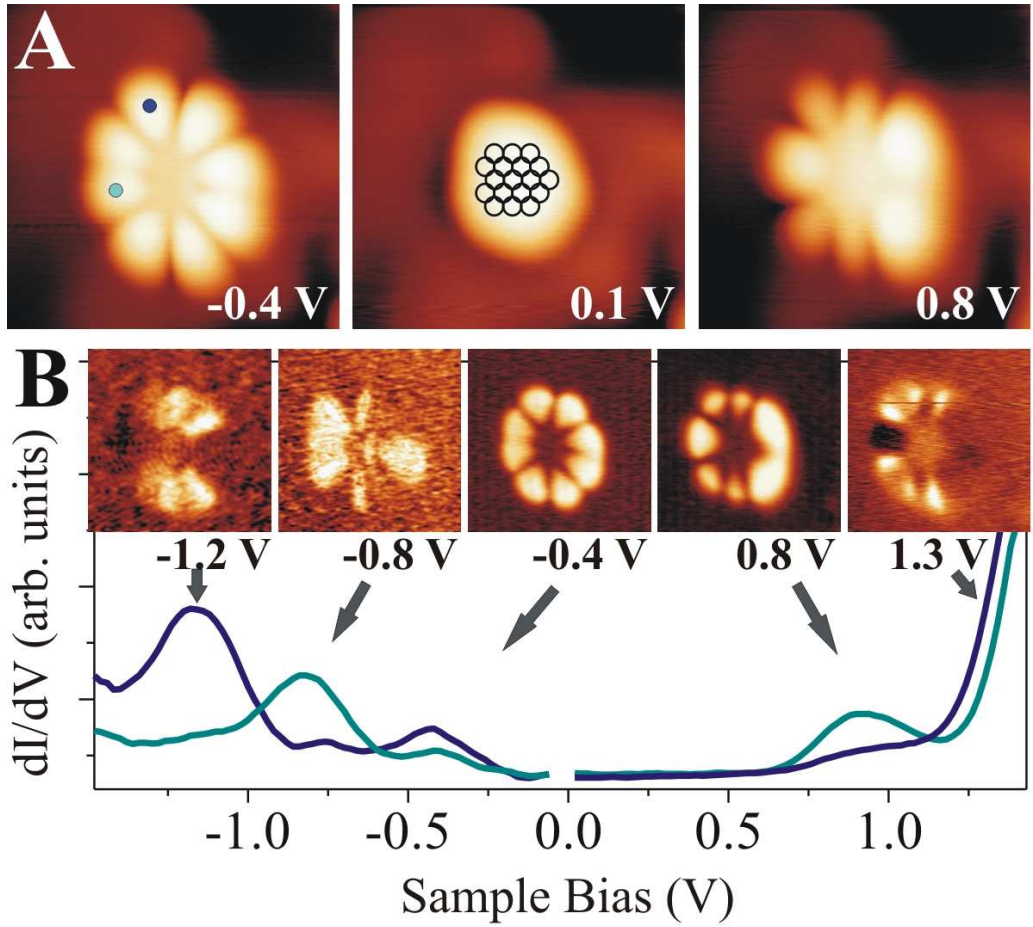

**Fig. 2**
**(A)** Topographic and **(B)** dI/dV images of a symmetric Au cluster on 2 ML MgO/Ag(001) taken at the given sample bias (I = 5 pA, 3.9×3.9 nm$^2$). The corresponding dI/dV spectra are shown in addition (blue and cyan curves: top and left part of the cluster). The structure model in (A) is deduced from the DFT.

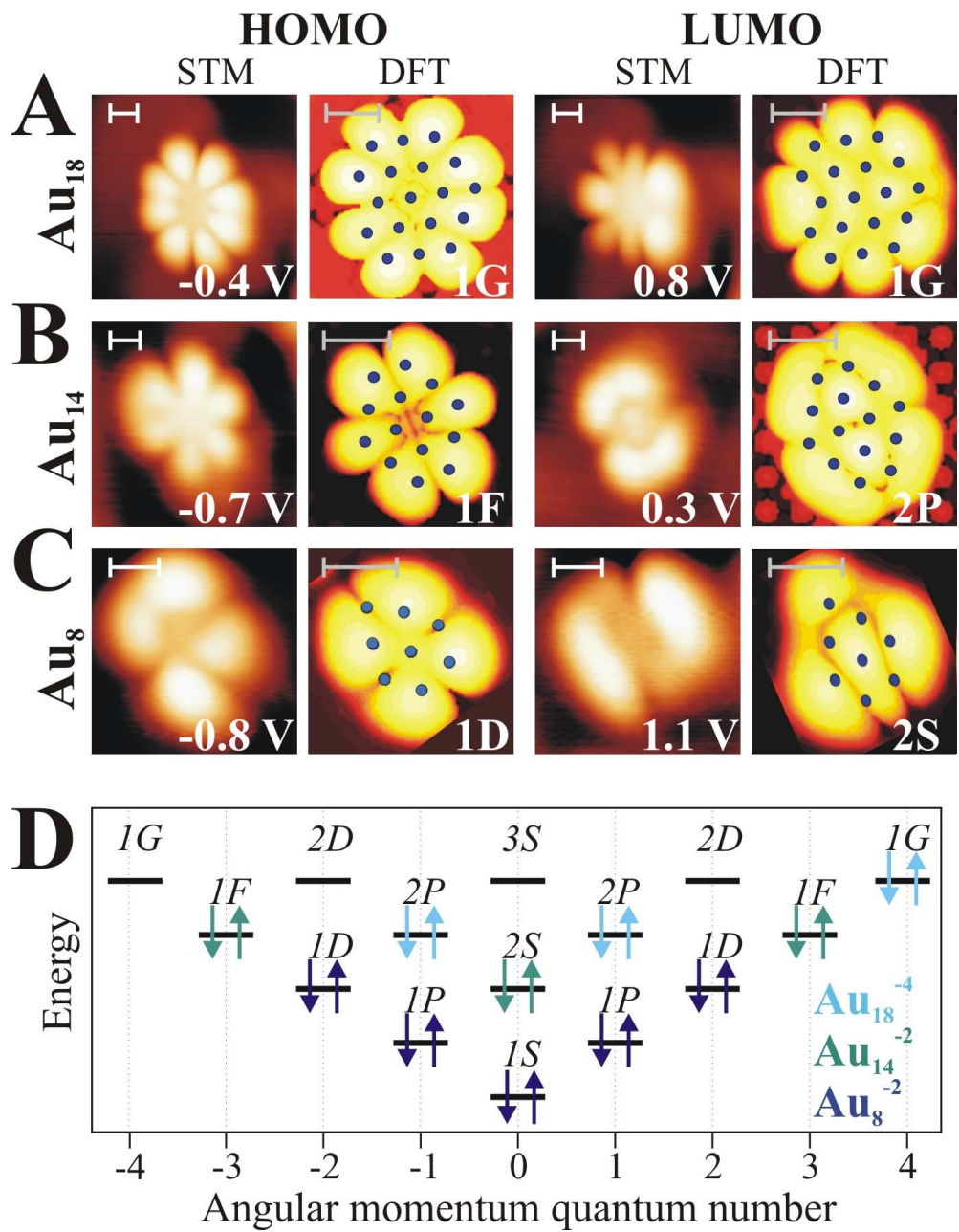

**Fig. 3**
**(A-C)** Measured and calculated HOMO and LUMO shapes for three differently sized Au clusters on MgO thin films. The dots in the calculated images depict the Au positions. All scale bars are 5 Å. **(D)** Orbital filling of the three clusters in the harmonic oscillator model. The $Au_8$ carries eight intrinsic plus two transfer electrons from the support. The $Au_{14}$ and $Au_{18}$ clusters are two- and fourfold negatively charged, respectively.

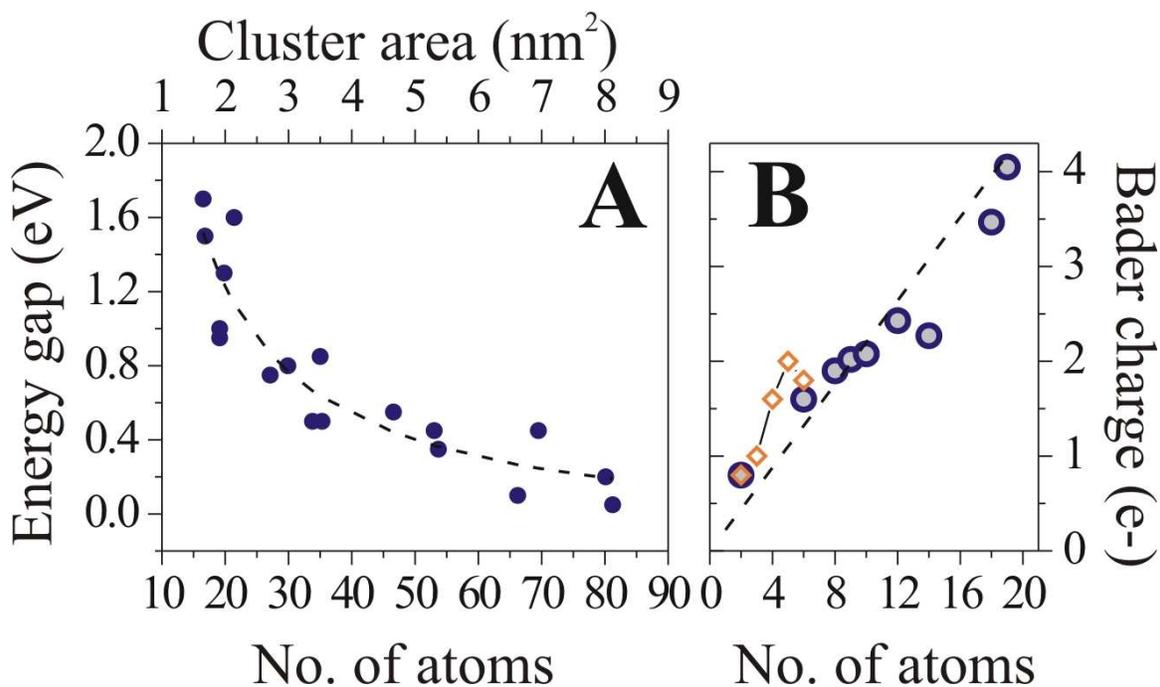

**Fig. 4**
(**A**) Experimental HOMO-LUMO gap for different Au clusters. The atom number is approximated from de-convoluted STM images, using the DFT results as a reference. The dashed line is a fit of the data to the inverse cluster size. (**B**) Calculated number of excess electrons in linear (orange)[27] and 2D Au clusters (blue) on MgO thin films. The line depicts the average charging per atom in a compact Au layer on 2 ML MgO/Ag(001).